\documentclass[11pt,english]{article}
\usepackage[T1]{fontenc}
\usepackage[utf8]{luainputenc}
\usepackage{amsmath}
\usepackage{amssymb}
\usepackage{setspace}
\usepackage{esint}
\makeatletter
\usepackage{babel}

\makeatother

\usepackage{babel}
\begin{document}

\title{The spacetime DoF surface density as canonically conjugate variables
driven by acceleration}

\author{Merav Hadad{\small{}{} }\thanks{{\small{}{}meravha@openu.ac.il }}}
\maketitle
\begin{abstract}
Padmanabhan found a definition for the surface density of spacetime
degrees of freedom assuming a canonical ensemble for accelerating
observers in stationary systems. We find that this density is constructed
from a pair of canonically conjugate variables. These variables should
be derived with respect to a special space like vector field, which
is in the direction of the accelerating observers. Moreover, we show
that this density is actually a pair of canonically conjugate variables
projected along the observer's velocity vector field, during one period
of the Euclidean time of the canonical ensemble. We use our derivation
to estimate the number of internal states of the \textquotedbl{}atoms
of spacetime\textquotedbl{}. We find that this number is independent
on the gravitational theory and its dimensions.
\end{abstract}
It is well known that an inertial detector will not always register
the same particle density as an accelerating detector and a uniformly
accelerated detector will 'see' thermal radiation\cite{Davies (1975),Unruh (1976)}
even though the field is in the vacuum Minkowski state. Padmanabhan
\cite{Padmanabhan:2010xh} used the Davies-Unruh temperature for accelerating
observers and found that the surface density of spacetime degrees
of freedom is given by 
\begin{eqnarray}
32\pi U_{cd}^{ab}\epsilon_{ab}\epsilon^{cd}\label{canonical string-1}
\end{eqnarray}
where $U^{abcd}\equiv\frac{\partial L}{\partial R_{abcd}}$ and $\epsilon_{ab}$
is the bi-normal to the surface. In this paper we suggest that Padmanabhan's
surface density can be constructed from a pair of special canonically
conjugate variables of accelerating reference frames.

The idea that the space-time degrees of freedom may be constructed
from a pair of canonically conjugate variables is not new. Carlip
and Teitelboim \cite{Carlip:1993sa} found that the Euclidean opening
angle is canonically conjugate to the local measure of area on the
horizon of the black hole and summing this over all horizon geometries
yields the black hole entropy. This property was extended to black
holes in general theories of gravity \cite{Brustein:2012sa}.

However, Padmanabhan's surface density of spacetime degrees of freedom
was found by using the equipartition relation. He used the idea that
in thermal equilibrium each degree of freedom carries an amount of
energy $\frac{1}{2}k_{B}T$ and thus the ``total degrees of freedom''
will be the ``total energy'' divided by the energy of each degree
of freedom. Taking the total energy to be the Komar mass, the temperature
to be the Davies-Unruh temperature for an accelerating observer, and
using the gravitational field equations for $D$ dimensional Lanczos-Lovelock
models, Padmanabhan was able to establish the spacetime degrees of
freedom surface density. He found that this surface density depends
only on aspects of the gravitational theory and on the surface spanned
by the bi-normal $\epsilon_{ab}=\frac{1}{2}(n_{a}u_{b}-n_{b}u_{a})$
where $u^{a}$ and $n^{a}$ are unit vectors of the velocity and acceleration,
respectively. Padmanabhan also suggested that this surface density
of spacetime degrees of freedom may be some kind of canonically conjugate
variables.

In this paper we argue that indeed Padmanabhan's density of spacetime
degrees of freedom can be constructed from a pair of special kind
of canonically conjugate variables. It is special since, instead of
deriving it with respect to some time like vector, it is derived with
respect to a space like vector. The fact that one may find a useful
spacetime canonically conjugate variables using space like, instead
of time like vector, for foliating spacetime, is not new \cite{Hadad:2015gtr}
and is even standard in the membrane paradigm. Moreover, we found
that the special space like vector field should be unique; it is the
direction of the accelerating vector field. We find that this foliation
leads to an interesting property: it enables us to distinguish between
the spacetime degrees of freedom detected by an accelerating detector,
from those detected by some free-falling detector at the same point.
This distinction is logical in light of the equivalence principle
combined with the Unruh effect. We show that projecting these canonically
conjugate variables along the velocity vector field, during one period
of the Euclidean time of the canonical ensemble, is, up to numerical
factors, exactly Padmanabhan's surface density of spacetime degrees
of freedom. Finally, we use our derivation to estimate the \textquotedbl{}atoms
of spacetime\textquotedbl{} internal state number observed by an accelerating
observer and find that this number is independent on the gravitational
theory and its dimensions.

We start by defining the direction of the space-like vector field
in a stationary D-dimensional spacetime. We take accelerating detectors
that have a $D$ velocity unit vector field $u^{a}$ and acceleration
$a^{a}=u^{b}\nabla_{b}u^{a}\equiv an^{a}$ (where $n^{a}$ is a unit
vector and $u^{a}n_{a}=0$)\footnote{We assume that both unit vectors: $n^{a}$and $u^{a}$ are hyper surface
orthogonal and thus fulfill Frobenius's theorem}. We limit our derivation to stationary metrics that are written in
coordinates system that cause the metric to be time independent. Moreover,
we limit the velocity vector field $u^{a}$ to represent observers
that \textquotedbl{}stand still\textquotedbl{} in these coordinate
system. Note that this causes the acceleration vector field $a^{a}$
to be time independent as well. Next we foliate spacetime with respect
to the unit vector field $n_{a}$ by defining a $(D-1)$- hyper-surface
which is normal to $n_{a}$. The lapse function $M$ and shift vector
$W_{a}$ satisfy $r_{a}=Mn_{a}+W_{a}$ where $r^{a}\nabla_{a}r=1$
and $r$ is constant on $\Sigma_{D-1}$. The $\Sigma_{D-1}$ hyper-surfaces
metric $h_{ab}$ is given by $g_{ab}=h_{ab}+n_{a}n_{b}$. The extrinsic
curvature of the hyper-surfaces is given by $K_{ab}=-\frac{1}{2}\mathcal{L}_{n}h_{ab}$
where $\mathcal{L}_{n}$ is the Lie derivative along $n^{a}$.

As was first noted by Brown \cite{Brown:1995su} for generalized theories
of gravity, the canonical conjugate variable of the extrinsic curvature
$K_{bc}$ is $4\sqrt{-h}n_{a}n_{d}U_{0}^{abcd}$. Where $U_{0}^{abcd}$
is an auxiliary variable, which equals $\frac{\partial\mathcal{L}}{\partial R_{abcd}}$
when the equations of motion hold. There are other canonical conjugate
pairs, such as $h_{ab}$ and its canonical conjugate variable. Some
may even depend on Brown's canonical pairs. However, surprisingly
we found that in order to express Padmanabhan's surface density as
canonically conjugate variables we do not need to find and verify
all the spacetime independent canonically conjugate variables.\footnote{We believe that this fact by itself is interesting and that these
canonically conjugate variables should be farther investigated since
it may be regarded as the relevant phase space of an accelerating
observer.} Note that since we limit our derivation to stationary metrics and
a velocity vector field which represent observers that \textquotedbl{}stand
still\textquotedbl{}, Brown's canonical conjugate variables do not
depend on time.

The next step is ``picking out'' the relevant canonically conjugate
variables that we expect to detect by the accelerating detectors.
We suggest to distinguish these canonically conjugate variables by
projecting the canonically conjugate tensors along the time like unit
vector $u_{b}$ which is the velocity of the accelerating detectors.
This suggestion is in agreement with different kind of physical quantities
which are calculated with respect to an observer's trajectory such
as energy densities $\rho=T_{ab}u^{a}u^{b}$ and Komar mass density
$\rho_{Komar}=\left(T_{ab}-\frac{1}{2}Tg_{ab}\right)u^{a}u^{b}$.
Thus we argue that the relevant conjugate variables for detectors
with D-velocity $u^{a}$ at point $P$ can be identified by projection
of the extrinsic curvature tensor $u^{a}$:\footnote{This means that we distinguish these canonically conjugate variables
from the others by projecting the extrinsic curvature and its canonical
conjugate variable along the time like unit vector $u_{b}$. Actually
this should be done more carefully since the Lie derivative of the
normal vector $u_{b}$ does not vanish in general and thus leads,
for example, to a contribution to the canonical conjugation of $h_{ab}$.}

\begin{eqnarray}
\left\{ K^{nm}u_{m}u_{n}(x),4\sqrt{h}U_{0}^{abcd}n_{a}u_{b}u_{c}n_{d}(x)\right\} 
\label{tensor canonical tern}
\end{eqnarray}
where we mark the coordinate by $(t,r,x)=(t,r,x_{1},...,x_{D-2})$.
Using $K^{ab}u_{b}u_{a}=n^{a}a_{a}=a$ we find: 
\begin{eqnarray}
\left\{ a(x),4\sqrt{h}U_{0}^{abcd}n_{a}u_{b}u_{c}n_{d}(x)\right\} 
\label{eq:general canonical term}
\end{eqnarray}

Before deriving the DoF let us evaluate this pair of special canonically
conjugate variables in two theories: Einstein theory where $\mathcal{L}=\frac{1}{16\pi G}R$
and for the gravitational theory $\mathcal{L}=CR_{abcd}R^{abcd}$
. For Einstein theory, using $U_{0}^{abcd}=\frac{\partial\mathcal{L}}{\partial R_{abcd}}=\frac{1}{16\pi G}\frac{1}{2}\left(g^{ac}g^{bd}-g^{ad}g^{bc}\right)$,
we find that $U_{0}^{abcd}n_{a}u_{b}u_{c}n_{d}=\frac{1}{32\pi G}\left(g^{ac}g^{bd}-g^{ad}g^{bc}\right)n_{a}u_{b}u_{c}n_{d}=\frac{1}{32\pi G}$.
Thus, according to our observation, the pair of conjugate variable
relevant to the gravitational DoF which is expected to be observed
by an accelerating detector in Einstein theory at point $P$ is
\begin{eqnarray}
\left\{ a(x),\frac{\sqrt{h(x)}}{8\pi G}\right\} 
\label{eq:Einst canonical term}
\end{eqnarray}
For the gravitational theory $\mathcal{L}=CR_{abcd}R^{abcd}$ , using
$U_{0}^{abcd}=\frac{\partial\mathcal{L}}{\partial R_{abcd}}=CR^{abcd}$,
we find that $U_{0}^{abcd}n_{a}u_{b}u_{c}n_{d}=CR^{abcd}n_{a}u_{b}u_{c}n_{d}=CR_{cd}^{ab}\epsilon_{ab}\epsilon^{cd}$
(where $\epsilon_{ab}=\frac{1}{2}(n_{a}u_{b}-n_{b}u_{a})$ ). Thus,
according to our observation, the pair of conjugate variable relevant
to the gravitational DoF which is expected to be observed by an accelerating
detector in the theory $\mathcal{L}=CR_{abcd}R^{abcd}$ at point $P$
is
\begin{eqnarray}
\left\{ a(x),4C\sqrt{h}R_{cd}^{ab}\epsilon_{ab}\epsilon^{cd}(x)\right\} 
\label{eq:exp. canonical term}
\end{eqnarray}
As expected, the first variable does not depend on the gravitational
theory. This fact turns to be important when comparing our derivation
to the membrane paradigm.

Next we turn to deriving the gravitational density degrees of freedom.
We argue that the gravitational density degrees of freedom detected
by an accelerating detector with D-velocity $u^{a}$ at point $P$
is constructed from multiplying these special stationary canonically
conjugate variables
\begin{eqnarray}
 &  & 4\sqrt{h}U_{0}^{abcd}a_{a}u_{b}u_{c}n_{d}(x)\label{canonical string1-1}
\end{eqnarray}

However, due to the foliation along a space like vector field, this
gravitational density is a density per $D-2$ area surface and time
(instead of volume density which is expected when foliating spacetime
with respect to time like vector field). This means that the number
of DoF detected by an accelerating detector grows with time. This
property seems to be a generalization of Unrue radiation which is
expected in flat spacetime,to any curved space time in any theory
of gravity. Thus, the gravitational $D-2$ surface density of the
spacetime DoF observed by an accelerating observer $\Delta n$ per
unit time $\Delta t$ is
\begin{eqnarray}
\frac{\Delta n}{\Delta t} & = & 4\sqrt{h}U_{0}^{abcd}a_{a}u_{b}u_{c}n_{d}(x).\label{density per time}
\end{eqnarray}
Moreover, note the dependence on the acceleration vector $n_{a}$:
it vanishes not only when the magnitude of the acceleration vanishes
as expected, but also if $u^{a}$ is a geodesic with a non affine
parameter. (In the geodesic with a non affine parameter $a^{a}=u^{b}\nabla_{b}u^{a}\equiv au^{a}$
and thus $a\neq0$ and so it seems that the canonical variables do
not vanish. However since in this case $n^{a}=u^{a}$, the term $4\sqrt{h}U_{0}^{abcd}n_{a}u_{b}u_{c}n_{d}$
vanishes from symmetry). This means that foliating spacetime with
respect to direction of the acceleration detectors enables us to distinguish
between the gravitational degrees of freedom which are detected by
the accelerating detectors from those detected by some free-falling
detectors at the same point. This unique foliation ``picks up''
a special direction which, by definition, is normal to the observer
velocity. All this suggests that these canonically conjugate variables
represents some extra degrees of freedom that detectors detect due
to their acceleration.

Finally, we derive $\Delta n$ by integrating this term during some
natural period of time. In order to find this period we use the basic
assumptions in \cite{Padmanabhan:2010xh}: an accelerated observer
detects her environment as an equilibrium system at temperature $T$
even in curved spacetime. This kind of canonical ensemble gives thermal
Greens functions which are periodic in the Euclidean time with period
$\beta=1/T$ \cite{B=000026D}. Assuming this is also the case for
fields which represent the spacetime degrees of freedom we integrate
eq. (\ref{density per time}) over the Euclidian period $\beta$\footnote{Since eq. (\ref{density per time})does not depend on time it does
not change sign in the range of integration and thus is expected to
be $L^{1}$ integrable.} and find that the surface density of the spacetime degrees of freedom:
\begin{eqnarray}
\Delta n=4\beta\sqrt{h}U_{0}^{abcd}a_{a}u_{b}u_{c}n_{d}(x).\label{density+beta}
\end{eqnarray}
Next, using the assumption that the temperature $T=1/\beta$ seen
by an accelerated observer equals $Na/2\pi$ even in curved spacetime
\cite{Padmanabhan:2010xh}, and $\sqrt{h}=N\sqrt{\sigma}$ ,we deduce
that the degrees of freedom surface density in a one period Euclidean
time detected by an accelerating observer is: 

\begin{eqnarray}
\Delta n=8\pi\sqrt{\sigma}U_{0}^{abcd}n_{a}u_{b}u_{c}n_{d}(x).\label{density}
\end{eqnarray}
As example, for Einstein theory we find that 
\begin{eqnarray}
 &  & \Delta n=\frac{\sqrt{\sigma}}{4G}\label{Einst density}
\end{eqnarray}
with is the expected result up to factor 4 and for the theory $\mathcal{L}=CR_{abcd}R^{abcd}$:
\begin{eqnarray}
 &  & \Delta n=8\pi CR^{abcd}n_{a}u_{b}u_{c}n_{d}(x)\label{exap density}
\end{eqnarray}
For a general modified theory, using the surface spanned by the bi-normal
$\epsilon_{ab}=\frac{1}{2}(n_{a}u_{b}-n_{b}u_{a})$ and $\mathbf{\boldsymbol{\mathbf{\epsilon}}_{ab}}=\sqrt{\sigma}\epsilon_{ab}$
we find that 
\begin{eqnarray}
 &  & \Delta n=8\pi U_{0}^{abcd}\epsilon_{a}{}_{b}\mathbf{\mathbf{\mathbf{\boldsymbol{\epsilon}}}_{c}{}_{d}}(x).\label{epsilon density}
\end{eqnarray}
Up to factor of 4, this is precisely the expression for entropy density
of spacetime found by Padmanabhan using the equipartition law for
a static metric. Thus, we found that the spatioal foliation provides
a way to evaluate the extra DoF detected by observers accelerating
in the direction of the foliation.

It turns out that this unique foliation is standard in the membrane
paradigm. The membrane paradigm models a black hole as a thin, classically
radiating membrane vanishingly close to the black hole's event horizon.
We note that the direction being used in deriving the membrane is
the radial direction which is the direction of the acceleration of
an observer hovering outside the black hole. Our derivation suggests
that the radiation coming from the modeled membrane is uniquely related
to foliating spacetime in the direction of an obverse's acceleration.
Moreover, since in Einstein theory, the membrane energy density is
the $u-u$ component of the membrane stress tensor $S_{ab}=K_{ab}-h_{ab}K$,
it is tempting to relates the $u-u$ component of the extrinsic curvature,
which is first the canonical part, to the energy density of the membrane.
However, as we noted in eq. (\ref{eq:general canonical term}) for
modified theories of gravity, the $u-u$ component of the first canonical
term (i.e. the extrinsic curvature) does not depend on the gravitational
theory, whereas the stress tensor does. Thus, in order to relate the
energy density of the membrane to the DoF density, we need to multiplys
the DoF density with the energy related to each DoF; i.e. half the
temperature. This leads to the heat density of the membrane \cite{Padmanabhan:2010xh}
for Lanczos-Lovelock theories.

Note that our derivation is expected to hold even for a time dependent
gravitational theories and even a time dependent accelerating observers.
In this case the only difference is that the gravitational $D-2$
surface density of the spacetime DoF observed by an accelerating observer
$\Delta n$ per unit time $\Delta t$ will be time dependent
\begin{eqnarray}
\frac{\Delta n(t,x)}{\Delta t} & = & 4\sqrt{h}U_{0}^{abcd}a_{a}u_{b}u_{c}n_{d}(t,x).\label{time dependent}
\end{eqnarray}
However in the time dependent case, we do not expect to be able to
describe the system as a stationary canonical ensemble. This means
that the theory is not periodic in Euclidian time, and thus we are
unable us to evaluate the area density of the gravitational DoF detected
by an accelerating observer during that time. We do expect the time
dependent density per area and time to hold.

The fact that the spacetime surface density observed by an accelerating
observer can be constructed from a pair of canonically conjugate variables
is interesting.This suggests, at least for accelerating observers,
that these variables are reasonable candidates for the phase space
of the spacetime microstates, in the same sense that position and
momentum are the phase space of matter microstates. Note that though
the inner structure of matter microstates: molecules, atoms, etc.
, are very complicated, they all share common phase space: position
and its canonical conjugate momentum. If we continue the analogy between
the matter microstates and the spacetime microstates, our result suggests,
at least for accelerating observers, that even if the unknown inner
structure of the spacetime microstates may be complicated, they probably
share common phase space: the extrinsic curvature of the surface directed
along the accelerating vector field and its canonical conjugate momentum.

Moreover, our observation can be used to estimate the number of the
internal states of the microscopic \textquotedbl{}atoms of spacetime\textquotedbl{}
observed by an accelerating observer. In order to do that we note
that if we attribute $f$ internal stats to each patch of phase space
$\Delta n$, then the total number of microstates $\Delta\varOmega$
will be $\Delta\varOmega=f^{(\Delta n)}$. Using the fact that the
entropy $\Delta S=ln(\Delta\Omega)=\Delta n\ln f$ , we find that
the number of internal state is expected to be $f=exp(\frac{\Delta s}{\Delta n}).$
For example, lets consider Einstein theory an observer hovering near
a Schwarzschild black hole horizon . Since this observer is actually
accelerating in the radial direction, we can use our derivation for
$\Delta n$ for Einstein theory $\Delta n=\frac{\sqrt{\sigma}}{4G}$.
Using the black hole entropy per unit area $\Delta s=\frac{\sqrt{\sigma}}{G}$,
we find the number of internal state of spacetime microstates is Einstein
theory is $f=exp(4)$. For modified theories of gravity, we use eq.
(\ref{exap density}) . Using Noether charge entropy for black holes
in general theories of gravity leads to the same conclusion: $f=exp(4)$.
Actually, using Padmanabhan's derivation for spacetime entropy area
density detected by an accelerating observer we fined the same result:
the number of internal state of spacetime $f=exp(4)$ for Lanczos-Lovelock
theories. Of course, the result $f=exp(4)$ may not be exclusive\footnote{It depends on our definition for $\Delta n$ . Why did we integrated
$\frac{\Delta n}{\Delta t}$ during ONE period of the Euclidean time
of the canonical ensemble? If the integration is over a quarter of
the periodic time we find that $\Delta n=\Delta s$ . In this case
$f=exp(1)=e$. }. However, the conclusion that, if indeed space-time microstates exist,
there internal number of state does not depend nor on the gravitational
theory and nor on the number of dimension is expected hold.

To summarize, in this paper we derived a special canonically conjugate
variables by foliating spacetime with respect to the direction of
an accelerating vector field. We found that the projection of these
canonically conjugate variables along the velocity vector of accelerating
detectors vanishes if the magnitude of the acceleration is zero or
even for a geodesic with a non affine parameter. Evaluating these
canonically conjugate variables, during one period of periodic Euclidean
time of the canonical ensemble expected for accelerating observers
gives the same surface density of spacetime degrees of freedom found
by Padmanabhan. This result proves that the gravitational surface
density can be constructed from a pair of special kind of canonically
conjugate variables if one foliates spacetime with respect to the
direction of the accelerating vector field. This unique foliation
enables us to distinguish between the spacetime degrees of freedom
which are detected by an accelerating detector, and the one detected
by some freely-falling detector at the same point. Moreover, our derivation
can be extended for time dependent theories and can be used to estimate
the number of the internal states of the microscopic \textquotedbl{}atoms
of spacetime\textquotedbl{} detected by an accelerating observer.
We find that this number is independent on the gravitational theory
and its dimensions.

\end{document}